\documentclass[twocolumn,twocolappendix]{aastex631} %linenumbers twocolumn

\usepackage[utf8]{inputenc}
\usepackage{enumerate}
\usepackage{graphicx}
\usepackage{mathtools}
\usepackage{comment}
\usepackage{tabularx}
\usepackage{amsmath}
\usepackage{changepage}
\usepackage{amssymb}
\usepackage{lipsum} 
\usepackage{float}
\usepackage{commath}
\usepackage{enumerate}
\usepackage{longtable}
\usepackage{stackengine}
\usepackage{xcolor}
\usepackage{cleveref}
\usepackage[caption=false]{subfig}
\usepackage{placeins}
\usepackage[ruled,lined]{algorithm2e}
\usepackage{hyperref}

\AfterEndEnvironment{figure}{\noindent\ignorespaces}

\begin{document}
\title{Galactic Accelerations from the GD-1 Stream Suggest a Tilted Dark Matter Halo} 
\author[0000-0001-8042-5794]{Jacob Nibauer}
\altaffiliation{NSF Graduate Research Fellow}
\affiliation{Department of Astrophysical Sciences, Princeton University, 4 Ivy Ln, Princeton, NJ 08544, USA}
\author[0000-0002-7846-9787]{Ana Bonaca}
\affiliation{The Observatories of the Carnegie Institution for Science, 813 Santa Barbara Street, Pasadena, CA 91101, USA}

\correspondingauthor{Jacob Nibauer}
\email{jnibauer@princeton.edu}
%TC:ignore
\begin{abstract}
\noindent
Cold dark matter halos are expected to be triaxial and often tilted relative to the stellar disk. Stellar streams provide a sensitive tracer of the Milky Way's halo shape, though models for the Galactic potential are typically limited to simple, symmetric functional forms. Here, we measure the Galactic acceleration field along the GD-1 stellar stream using a direct differentiation of the stream's track in phase-space. Using a fully data-driven catalog of stream members from {\it Gaia}, SDSS, LAMOST, and DESI, we map the stream in 6D phase-space. We fit splines to the stream track, and infer cylindrical acceleration components $a_R = -2.5 \pm_{0.1}^{0.2}, \ a_z = -1.8\pm 0.1, \ a_\phi = 0.2\pm 0.1~\rm{km \ s^{-1} \  Myr^{-1}}$ at $(R,z,\phi) = (11.9~\rm{kpc}, 7.3~\rm{kpc}, 171.1~\rm{deg})$. 
We measure mass enclosed within $14~\rm{kpc}$ of $1.4\pm 0.1 \times 10^{11} M_\odot$ and z-axis density flattening of $q_{\rho, z} = 0.81\pm^{0.06}_{0.03}$, both consistent with previous estimates. However, we find a 2$\sigma$ deviation from an axisymmetric acceleration field, which can be explained by a triaxial dark matter halo with axis ratios 1:0.75:0.70. The major axis of the halo is consistent with a tilt of $18~\rm{deg}$ above the Galactic plane in the direction of the Sun. The magnitude and direction of the tilt are consistent with measurements of the Milky Way's stellar halo from {\it Gaia} and the H3 survey. A tilted triaxial halo has important consequences for orbit-integration-based studies of the Galaxy, and can be further tested by deriving acceleration constraints from multiple streams.
\vspace{1cm}
\end{abstract}
%TC:endignore

\section{Introduction}
The distribution of matter in the Milky Way is of fundamental importance for studies of the Galaxy. However, there is a lack of consensus on the shape of the Milky Way's dark matter halo. Cosmologically, cold dark matter halos are expected to be triaxial, and aligned with filamentary structure as a result of ongoing mergers \citep{1988ApJ...327..507F,1991ApJ...378..496D,1992ApJ...399..405W,1996MNRAS.281..716C,2002ApJ...574..538J,2005ApJ...627..647B,2006MNRAS.367.1781A,2011MNRAS.416.1377V,2012JCAP...05..030S,2016MNRAS.462.2668T,2023A&A...669A.132P}. Observationally, the Tully-Fisher relation (e.g., \citealt{1992ApJ...392L..47F}), lensing (e.g., \citealt{2006MNRAS.370.1008M,2009ApJ...695.1446E}), rotation curves (e.g., \citealt{2023PhRvD.107h3524B}), and tidal streams (e.g., \citealt{2010ApJ...714..229L, 2019MNRAS.486.2995M, 2023ApJ...954..195N}) imply that halo shapes can be significantly aspherical. Additionally, the shape of halos depend on the nature of dark matter, specifically within the inner $\sim 20~\rm{kpc}$ where the dark matter density is highest \citep{2000PhRvL..84.3760S,2000ApJ...535L.103Y,2001ApJ...547..574D,2002ApJ...564...60M,2013MNRAS.430..105P, 2022MNRAS.516.4543D,2024ApJ...974..223A}.

The kinematically cold tidal tails of globular clusters trace the underlying mass distribution of the Galaxy. As stars are lost from the progenitor, they extend along a series of similar, albeit different orbits. The result is two tidal tails, one leading and the other trailing the cluster (see \citealt{2025NewAR.10001713B} for a recent review of stellar streams). The morphology and kinematics of tidal tails provide perhaps the most sensitive probe of the enclosed mass and shape of the dark matter halo (see, e.g., \citealt{1999ApJ...512L.109J, 2013MNRAS.436.2386L, 2015ApJ...799...28P, 2016ApJ...833...31B,2018ApJ...867..101B,2023ApJ...954..195N,2023MNRAS.521.4936K}).

There is no clear consensus on the shape of the Milky Way's inner dark matter halo (i.e., within $30~\rm{kpc}$) as inferred from stellar streams or other tracers. For example, the GD-1 stream and the tidal tails of Palomar 5 (Pal 5) are consistent with a spherical, or at most slightly oblate or prolate halo \citep{2016ApJ...833...31B,2019MNRAS.486.2995M,2021MNRAS.502.4170R}. Equilibrium modeling of the Galaxy's globular cluster population suggests a prolate halo \citep{2019A&A...621A..56P}, and the phase-mixed Helmi Streams imply a mildly triaxial inner halo \citep{2024A&A...691A.277W}. Additionally, the velocity dispersion and radial number density profile of stars within $10~\rm{kpc}$ are consistent with a prolate geometry \citep{2016MNRAS.460..329B}.

Cosmological simulations predict that dark matter halos are often triaxial and tilted relative to the stellar disk, a configuration that should imprint a dynamical signature on stellar tracers (e.g., \citealt{1989MNRAS.237..785O,1999ApJ...513L.107D,2021ApJ...913...36E,2023ApJ...957L..24H, 2024ApJ...969...55N}). Dynamical tracers of the halo are also affected by time-dependence in the potential, resulting from the Large Magellanic Cloud (LMC; e.g., \citealt{2023MNRAS.518..774L}) and, more generally, the cosmological growth of the halo (e.g., \citealt{1999MNRAS.307..495H,2015A&A...584A.120B}). Previous work modeling streams typically employ simple functional forms for the Galactic potential, which limits the range of possible measurements that can be obtained from the data. The adoption of simplified global halo models may explain at least some of the variance in halo shape measurements derived from different streams and tracers.

 In this work, we present a data-driven constraint on the Galactic acceleration field that is independent of user-defined functional forms for the potential. We use a direct differentiation of the GD-1 stream track in phase-space (\citealt{2022ApJ...940...22N}, hereafter N22) to infer the 3D Galactic acceleration field along the stream. We focus on GD-1 due to the large number of radial velocity and proper motion measurements that have become recently available for this stream. The central premise of the method from N22 is that while a cold stream does not trace a single stellar orbit, it is composed of a series of local orbit segments. Differential changes in position and velocity along each segment can be measured to extract an acceleration: 
 \begin{equation}\label{eq: acceleration}
    \boldsymbol{a}\left(\phi_1\right) = \frac{d\boldsymbol{v}}{d\phi_1}\left( \frac{d\boldsymbol{x}}{d\phi_1} \cdot \frac{d\boldsymbol{x}}{d\phi_1}\right)^{-1/2} \Vert \boldsymbol{v}\left(\phi_1\right)\Vert,
\end{equation}
where $\phi_1$ is a phase-angle along the stream's elongated axis, increasing in the direction of motion\footnote{If the stream is moving to -$\phi_1$, Eq.~\ref{eq: acceleration} picks up a minus sign.}, and $(\boldsymbol{x}, \boldsymbol{v}$) is the local position and velocity of the stream. The advantage of Eq.~\ref{eq: acceleration} is that the stream's position and kinematic track provides a direct constraint on the acceleration field without having to model the underlying potential. The method has been validated on simulated data in a triaxial halo, and successfully recovers halo shapes and mass profiles (for details, see N22).  

The paper is organized as follows. In \S\ref{sec: data} we describe the dataset and stream fitting procedure, in \S\ref{sec: results} we present our results on Galactic accelerations, in \S\ref{sec: interpretation} we fit mass models to the inferred accelerations, and in \S\ref{sec: summary_and_discuss} we summarize and conclude. Throughout this work, we assume a right-handed Galactocentric reference frame with the Sun at position $(-8.1~\rm{kpc}, 0, 20.8~\rm{pc})$ with velocity $v_\odot = (12.9, 245.6, 7.78)~\rm{km/s}$ (\texttt{Astropy} v4.0 parameters, \citealt{astropy:2022}).

\begin{figure}
\centering\includegraphics[scale=.45]{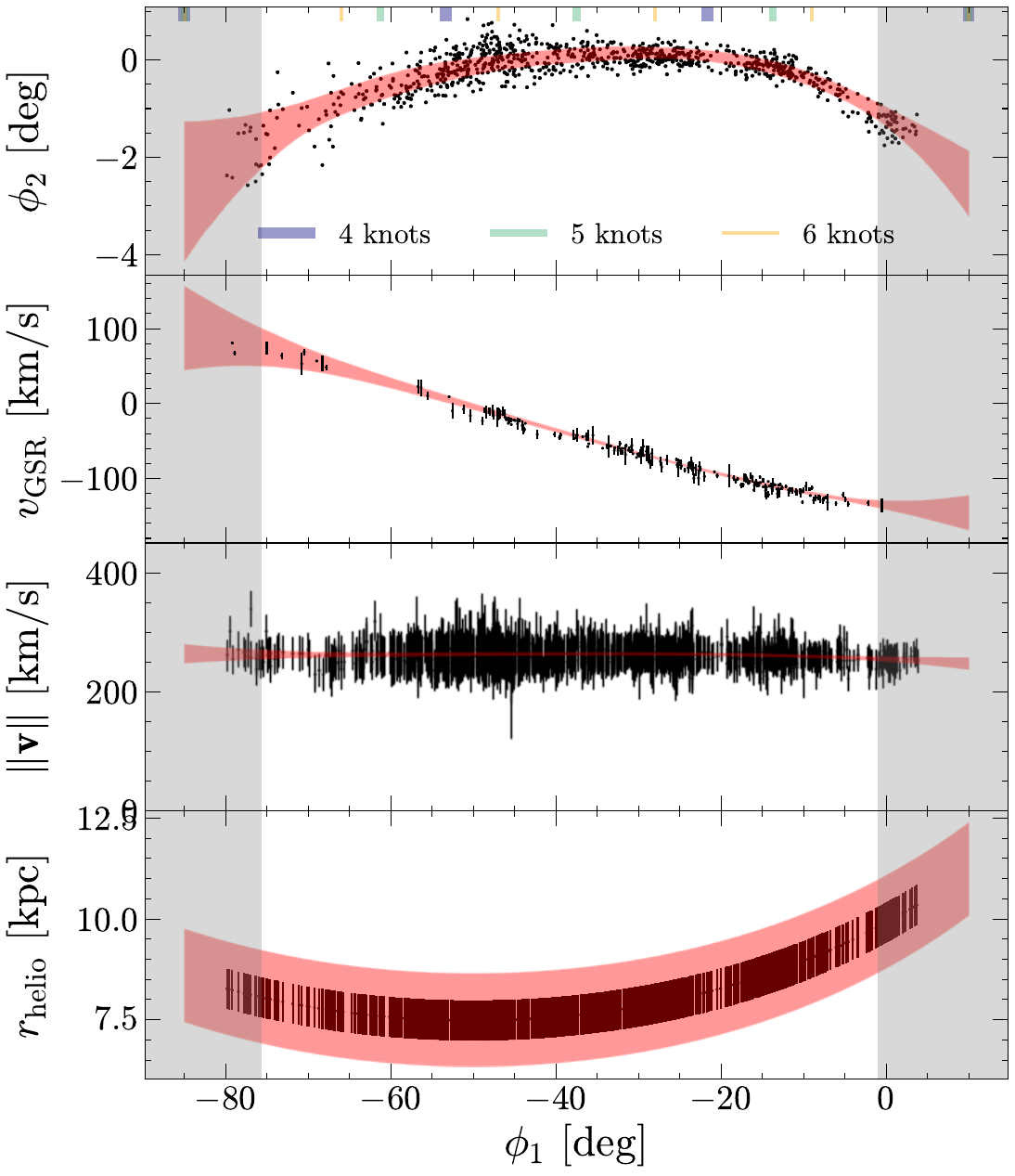}
    \caption{Black points indicate the data (with 1$\sigma$ errors), while the red band represents the 99\% credible interval of our stream-track for a spline with 5 knots. Knot locations are shown as lines in the top panel for the case of 4 (navy), 5 (green), and 6 (orange) knots. The $(\phi_1, \phi_2)$ data is from {\it Gaia} \citep{2025ApJ...980..253S}, solar-reflex correct radial velocities $v_{\rm GSR}$ are from DESI \citep{2025ApJ...980...71V}, SDSS 9 \citep{2012ApJS..203...21A}, and LAMOST DR8\textsuperscript{\ref{lamost}}. The distance track is from the sub-giant branch \citep{2025ApJ...980...71V}.}
    \label{fig: data}
\end{figure}

\section{Data and Stream Track Fitting}\label{sec: data}
We use a data-driven catalog of GD-1 stream-members that does not rely on a dynamical model. {\it Gaia} DR3 \citep{2023A&A...674A...1G} sky positions and proper motions of stream members are selected from \citet{2025ApJ...980..253S}. We select stars with membership probability $>65\%$, and only select stars from the main stream (no spur). Our conclusions do not change with more stringent membership cuts since we only model the stream track. Radial velocities are obtained from the Dark Energy Spectroscopic Instrument early data release (DESI; \citealt{2024AJ....168...58D,2024MNRAS.533.1012K, 2025ApJ...980...71V}), SDSS 9 \citep{2012ApJS..203...21A, 2019ApJ...877...13H}, LAMOST DR8\footnote{\url{http://www.lamost.org/dr8/} \label{lamost}}, and MMT Hectochelle \citep{2020ApJ...892L..37B}. For the radial velocity dataset, we select stream members within 10~\rm{km/s} of the mean radial velocity track from \citet{2025ApJ...980...71V}, and remove 3 stars whose total speed is $>5\sigma$ from the mean. The result is 910 stream members for the on-sky locations and proper motion measurements, and 245 members with radial velocities spanning a $\phi_1$ range of roughly 80~\rm{deg} on the sky. 175 of the radial velocity members are also identified as members based on sky positions and proper motions, while 65 are co-moving stars off the main stream track (from the spur and cocoon regions). Removing these off-track stars does not change our results, so we opted to keep them in our sample. {\it Gaia} parallaxes at the distance of GD-1 are highly uncertain. We estimate distances to GD-1 stream members by interpolating the distance track based on the subgiant branch \citep{2025ApJ...980...71V}, consistent with other dynamically measured distances \citep{2018ApJ...863L..20P,2018MNRAS.477.1893D,2018ApJ...869..122L}. We adopt systematic uncertainties of $0.5~\rm{kpc}$ for the distance track as the variance across literature models (see Fig. 13b of \citealt{2025ApJ...980...71V}). The dataset is illustrated in Fig.~\ref{fig: data}, where $\phi_1$ and $\phi_2$ are the on-sky angular coordinates of the stream \citep{2010ApJ...712..260K}. The radial velocity and total speed are solar-reflex corrected.

We use $C_2$ cubic splines to represent the track of GD-1 and infer accelerations using Eq.~\ref{eq: acceleration}. We use splines over the neural network approach in N22, because the track of GD-1 is sufficiently short and simple to be accurately captured with a simpler spline model. Splines are implemented using the package \texttt{Interpax} \citep{conlin_2025_14902770}. The on-sky location, radial velocity, and speed of particles (Fig.~\ref{fig: data}, black points) are easily captured by a spline with 4 knots. The minimum and maximum $\phi_1$ knot locations are $\phi_1 = -85~\rm{deg}$ and $\phi_1 = 5~\rm{deg}$, respectively. To ensure the robustness of our constraints, we marginalize over the number of knots by repeating our fitting process with $[4,5,6]$ evenly spaced knots in $\phi_1$ (marked as short lines at the top of Fig.~\ref{fig: data} in navy, green, and orange, respectively). The median inferred accelerations obtained for each choice of the number of knots are consistent within $1\sigma$. Because the speed of the stream as a function of $\phi_1$ is nearly flat, we fix the number of knots to $4$ when fitting $\Vert \mathbf{v}\Vert\left(\phi_1\right)$. We have tested a range of locations for the outermost knots, from the edges of the dataset to 15~\rm{deg} from each edge, and find consistent results on the derived accelerations.

We use the likelihood from N22 to fit splines to the data, and sample a range of tracks while accounting for observational uncertainties. We require the variance of the sampled tracks to reflect the intrinsic width and radial velocity scatter of the stream, and allow for possible misalignments between the stream's track and local velocity direction. This is different from how stream tracks are usually fit, since most studies characterize the local mean of the data so that the variance of fitted models is roughly $\sigma^2 / N$, where $\sigma^2$ is the local variance and $N$ is the local number of points in a $\phi_1$ bin. This approach gives a very narrow range of centerlines. To achieve a model variance of $\sigma^2 / N \xrightarrow{} \sigma^2$, we add (in quadrature) the estimated intrinsic scatter of the stream to the observational uncertainties, multiplying the intrinsic variance by the local number of data points ($N$) in roughly 4~\rm{deg} bins in $\phi_1$ for the $\phi_2(\phi_1)$ fit, and 8~\rm{deg} bins for the $v_{\rm GSR}(\phi_1)$ fit due to the lower number of data points. To estimate intrinsic scatter we use inverse variance weighting, accounting for observational uncertainties. Our inflation of errors is conservative, since this choice will produce less certain acceleration measurements. The sampling of many possible tracks within the width of the stream is consistent with N22, who allows tracks to deviate from the stream's centerline.

An example fit with 5 knots is shown in Fig.~\ref{fig: data} (red error bands). Sampling of the spline knots is performed in \texttt{JAX} \citep{jax2018github} using the NUTS sampler from \texttt{Blackjax} \citep{cabezas2024blackjax}. Following N22, we evaluate accelerations well within the edges of the data in $\phi_1$ (gray-shaded regions in Fig.~\ref{fig: data} are excluded), because derivatives are ill-defined at the edges.

\section{Results}\label{sec: results}
\begin{figure}
\centering\includegraphics[scale=.6]{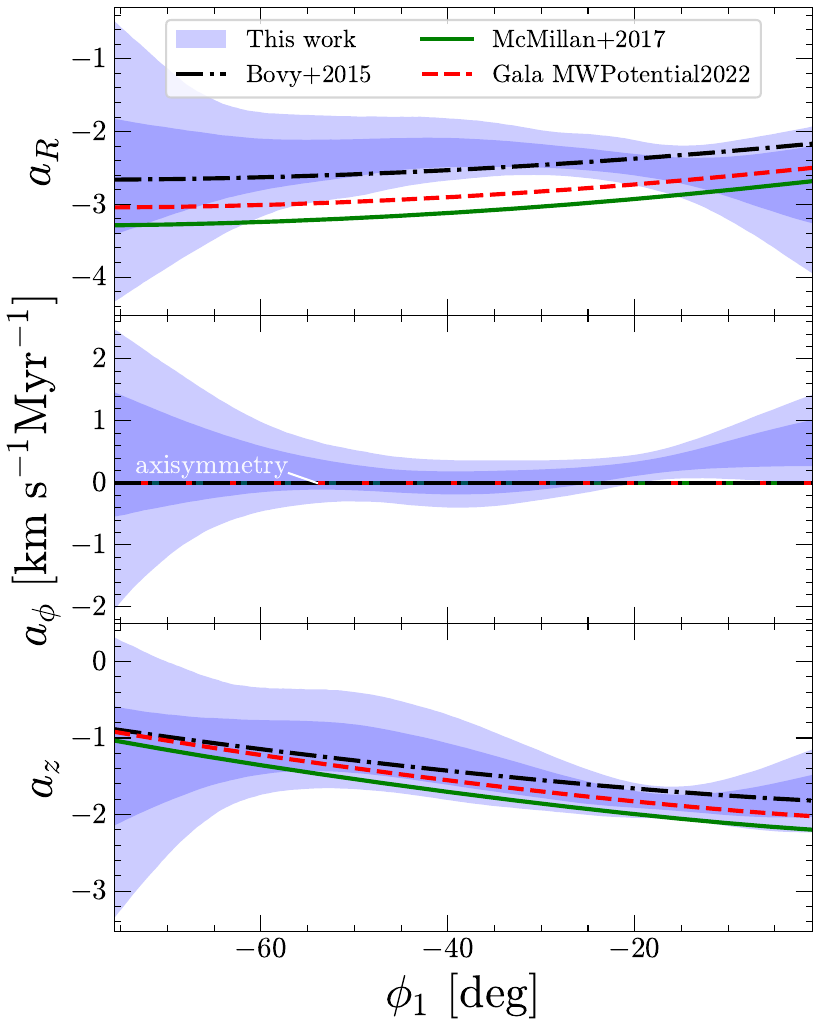}%analyze_production_withVariableDisk_andTestLMC_IP
    \caption{The inferred 3D accelerations from GD-1 in cylindrical Galacocentric components. Error-bands represent regions of 68 (dark) and 95\% (light) confidence.  Constraints are marginalized over 4-6 spline knots. We overplot three common potential models, including MWPotential2014 from \texttt{Galpy} (black; \citealt{2015ApJS..216...29B}), MilkyWayPotential2022 from \texttt{Gala} (red; \citealt{gala}), and the potential from \citet{2017MNRAS.465...76M} (green). The three models are axisymmetric ($a_\phi = 0$). Our data-driven accelerations are of a similar magnitude to all three models, though the best match is with \citet{2015ApJS..216...29B}. We find a 2$\sigma$ discrepancy from axisymmetry at $\phi_1 \approx -20~\rm{deg}$ where constraints are tightest.}
    \label{fig: Acceleration_Cylindrical}
\end{figure}

The inferred acceleration field along GD-1's track is shown in Fig.~\ref{fig: Acceleration_Cylindrical} in a Galactocentric cylindrical coordinate system. We use cylindrical coordinates because they can easily reveal deviations from axisymmetry ($a_\phi \neq 0$). Error bands represent regions of 68 and 95\% confidence, obtained by marginalizing over the number of spline knots, and sampling the posterior distribution of spline tracks (\S\ref{sec: data}). The inferred accelerations are in excellent agreement with MWPotential2014 from \texttt{Galpy} (black dot-dashed curve, \citealt{2015ApJS..216...29B}), though other common models also provide a reasonable match (solid green curve, \citealt{2017MNRAS.465...76M}; red dashed curve, \citealt{gala}). None of the three common potential models are fit to GD-1. Our accelerations are most precisely constrained around $\phi_1 = -15~\rm{deg}$, corresponding to a Galactocentric position $(R,z,\phi) = (11.9~\rm{kpc}, 7.3~\rm{kpc}, 171.1~\rm{deg})$, where all reported quantities are posterior medians. The acceleration components are
\begin{equation}
    \begin{split}
    a_R &= -2.5 \pm_{0.1}^{0.2}~\rm{km \ s^{-1} \  Myr^{-1}} \\
    a_z &= -1.8\pm 0.1~\rm{km \ s^{-1} \  Myr^{-1}} \\ 
    a_\phi &= 0.2\pm 0.1~\rm{km \ s^{-1} \  Myr^{-1}}.
    \end{split}
\end{equation}
Interestingly, our measurement indicates a 2$\sigma$ deviation from an axisymmetric mass distribution at this location, though the accelerations are consistent with axisymmetry ($a_\phi= 0$) at $\phi_1 \lesssim-20~\rm{deg}$ where the uncertainties are larger. 

\begin{figure}
\centering\includegraphics[scale=.42]{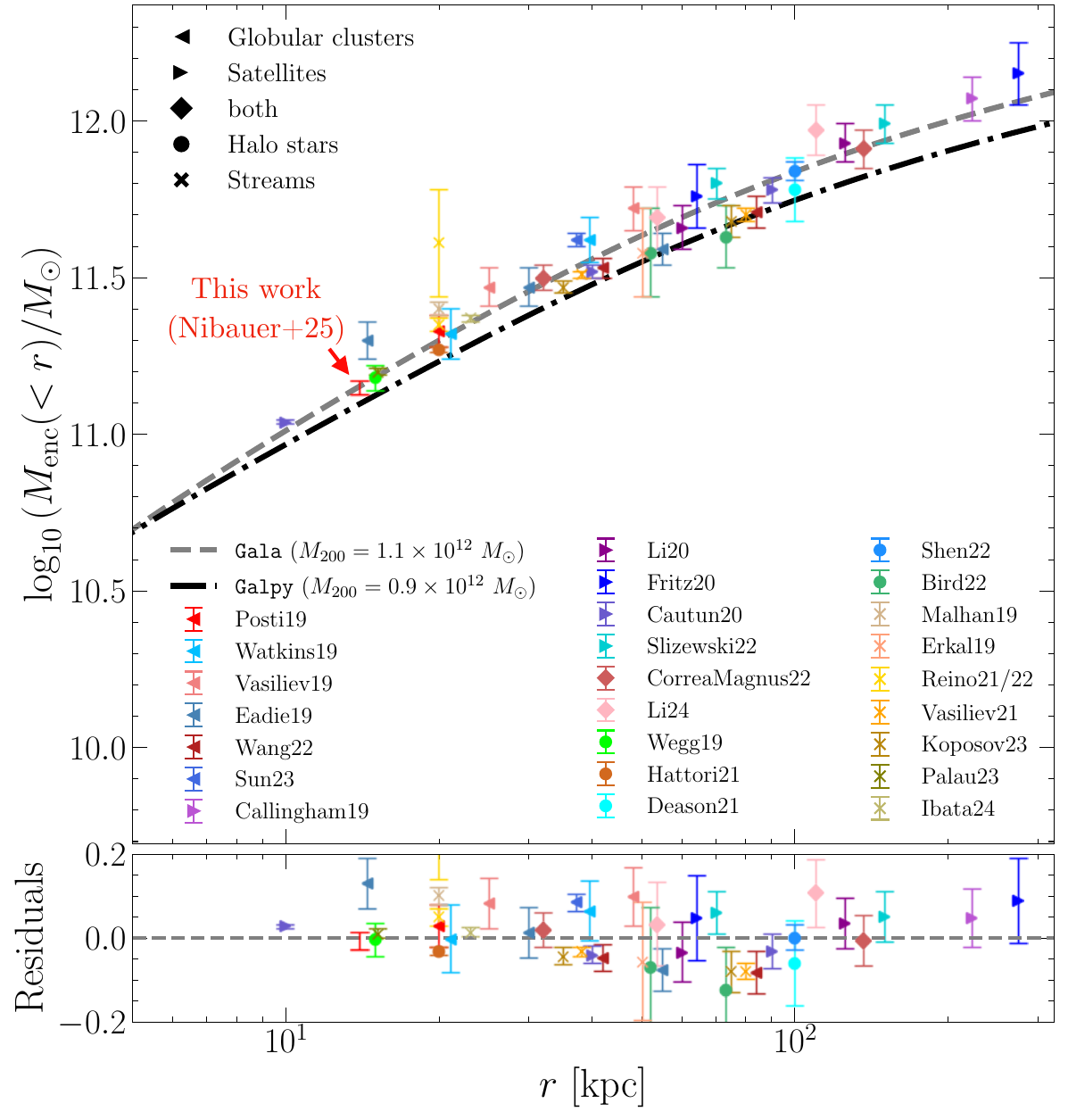}
    \caption{Mass-enclosed profile of the Milky Way, with data compiled from \citet{2025arXiv250104075H}. Our measurement is shown in red, below the labeled arrow, and agrees well with other inner-halo measurements. The tracer-type is indicated by the symbol, and the gray dashed gray line is MilkyWayPotential2022 from \texttt{Gala} \citep{gala}, while the dot-dash black line is MWPotential2014 from \texttt{Galpy} \citep{2015ApJS..216...29B}. Residuals are relative to MilkyWayPotential2022. See Appendix~\ref{app: fig_refs} for references. }
    \label{fig: Menc}
\end{figure}

The enclosed mass at $\phi_1 = -15~\rm{deg}$, approximated by $M_{\rm enc}(<r) = a_r r^2 / G$, is set by the amplitude of the $R$ and $z$ acceleration components. We find
\begin{equation}\label{eq: M_enc}
    M_{\rm enc}\left(r<14~\rm{kpc}\right) = \left(1.41\pm 0.07 \right)\times10^{11}~M_\odot 
\end{equation}
at 68\% confidence. In Fig.~\ref{fig: Menc} we compare our constraint to standard models of the Milky Way, and individual measurements from a recent compilation by \citet{2025arXiv250104075H}. Our $M_{\rm enc}$ measurement is consistent with MilkyWayPotential2022 ($1.43\times 10^{11}~M_\odot$; dashed line, \citealt{gala}), though higher than MWPotential2014 ($1.25\times 10^{11}~M_\odot$; dot-dashed line, \citealt{2015ApJS..216...29B}). While MWPotential2014 provides the best match to our accelerations across the length of the stream in Fig.~\ref{fig: Acceleration_Cylindrical}, near $\phi_1 = -15~\rm{deg}$ where our accelerations are most certain the MilkyWayPotential2022 model provides a better match to our inference. We also find good agreement with other works, particularly \citet{2019MNRAS.485.3296W} who used non-parametric Jeans modeling to infer the azimuthally averaged acceleration field of the Milky Way, and reports $M_{\rm enc}(r<15~\rm{kpc}) = 1.5\pm^{0.1}_{0.2} \times 10^{11}~M_\odot$. Our measurement of $M_{\rm enc}$ is also consistent with models of Pal 5's tidal tails \citep{2015ApJ...803...80K}.

Next, we compare the axisymmetric components of the inferred accelerations ($a_R$ and $a_z$) with the results from \citet{2016ApJ...833...31B}, the only other work that reports constraints on $a_R$ and $a_z$ from the GD-1 stream. \citet{2016ApJ...833...31B} assumes an axisymmetric mass model, while our method does not assume a mass model. Near the center of the stream we find the accelerations $a_R = -2.3\pm 0.2~\rm{km \ s^{-1} Myr^{-1}}, \ a_z = -1.4\pm^{0.3}_{0.2}~\rm{km \ s^{-1} Myr^{-1}}$ at $\phi_1 = -37~\rm{deg}$. For the same central region,  \citet{2016ApJ...833...31B} finds $a_R = -2.5\pm0.2~\rm{km \ s^{-1} Myr^{-1}}$ and $a_z = -1.4\pm 0.2~\rm{km \ s^{-1} Myr^{-1}}$. These values are consistent with our findings, providing a strong validation of the cylindrical $R$ and $z$ acceleration components near GD-1's location, despite the different models and assumptions used in our study and in \citet{2016ApJ...833...31B}.

We now measure the total flattening in the potential from the inferred accelerations, assuming the potential is locally of the form $\Phi\left(R^2 + (z/q_\Phi)^2 \right)$, where $q_\Phi$ is the total z-axis potential flattening. We can compute the flattening parameter for this potential using the expression
\begin{equation}
    q^2_\Phi = \frac{z}{R}\frac{a_R}{a_z}.
\end{equation}
We find $q_{\Phi} = 0.91 \pm 0.05$ at 68\% confidence where accelerations are most confident. This is consistent with $q_\Phi = 0.95 \pm 0.04$ from \citet{2016ApJ...833...31B}, and $q_\Phi = 0.87\pm^{0.07}_{0.04}$ from an orbit-fitting analysis of the GD-1 stream \citep{2010ApJ...712..260K}. Encouragingly, despite their different methodologies, all GD-1 studies agree that the Milky Way system is oblate.

\section{Mass Models Inferred from the GD-1 Accelerations}\label{sec: interpretation}
We now fit the accelerations inferred from GD-1 with global mass models for the Galaxy in order to constrain the global halo shape. We consider two models for the dark matter halo. The first consists of an axisymmetric halo with $z$-axis flattening perpendicular to the plane of the disk (\S\ref{sec: axisymm_halo}), and the second is a general triaxial halo oriented in an arbitrary direction (\S\ref{sec: triaxial_halo}).

The halo model is a Navarro–Frenk–White profile \citep{1997ApJ...490..493N}:
\begin{equation}\label{eq: G_NFW}
    \rho\left(\boldsymbol{x}^\prime\right) = \frac{M}{4\pi  r_s^3} \frac{1}{\left(m/r_s\right)\left(1 + m/r_s\right)^2},
\end{equation}
where $m^2 = x^{\prime 2} + \frac{y^{\prime 2}}{\left(b/a\right)^2} + \frac{z^{\prime 2}}{\left(c/a\right)^2}$, and $b/a, \  c/a$ are the $y^\prime$ and $z^\prime$ axis flattening values, respectively. In \S\ref{sec: axisymm_halo} the primed axes coincide with the Galactocentric axes $(x,y,z)$. In \S\ref{sec: triaxial_halo} we allow for the axes $x^\prime, y^\prime, z^\prime$ to rotate with respect to the Galactocentric axes $x,y,z$, and parametrize the rotation using a pitch angle (the angle between the Galactocentric $x-y$ plane and the semi-major axis) and a yaw angle (the azimuthal location of the semi-major axis from the Galactocentric $x$-axis). We omit fitting the third angle, roll, because it is not needed to describe GD-1 accelerations at the $1\sigma$ level. The halo scale-radius, $r_s$, is a free parameter in our fits ranging from 10-30~\rm{kpc}. For the baryonic components of the potential, we use the disk and spherical bulge from MWPotential2014 \citep{2015ApJS..216...29B}.

To fit a global halo model, we compare the accelerations predicted by a mass model to the measured accelerations along GD-1's median position track. We follow N22, and assume Gaussian uncertainties on each acceleration measurement, with standard deviations derived from the posterior variance of the inferred accelerations along the stream. We use \texttt{AGAMA} \citep{2019MNRAS.482.1525V} to derive a potential model from the density function, Eq.~\ref{eq: G_NFW}. Sampling is performed using \texttt{emcee} \citep{2013PASP..125..306F}

\subsection{Axisymmetric Halo Aligned with the Disk}\label{sec: axisymm_halo}
We now fit an axisymmetric model to the accelerations from our analysis. Free parameters include the halo mass $M$, the $z-$axis halo flattening $q_{\rho_z} \equiv c/a$, scale-radius $r_s$, and the disk mass. 

\begin{figure}
\centering\includegraphics[scale=.43]{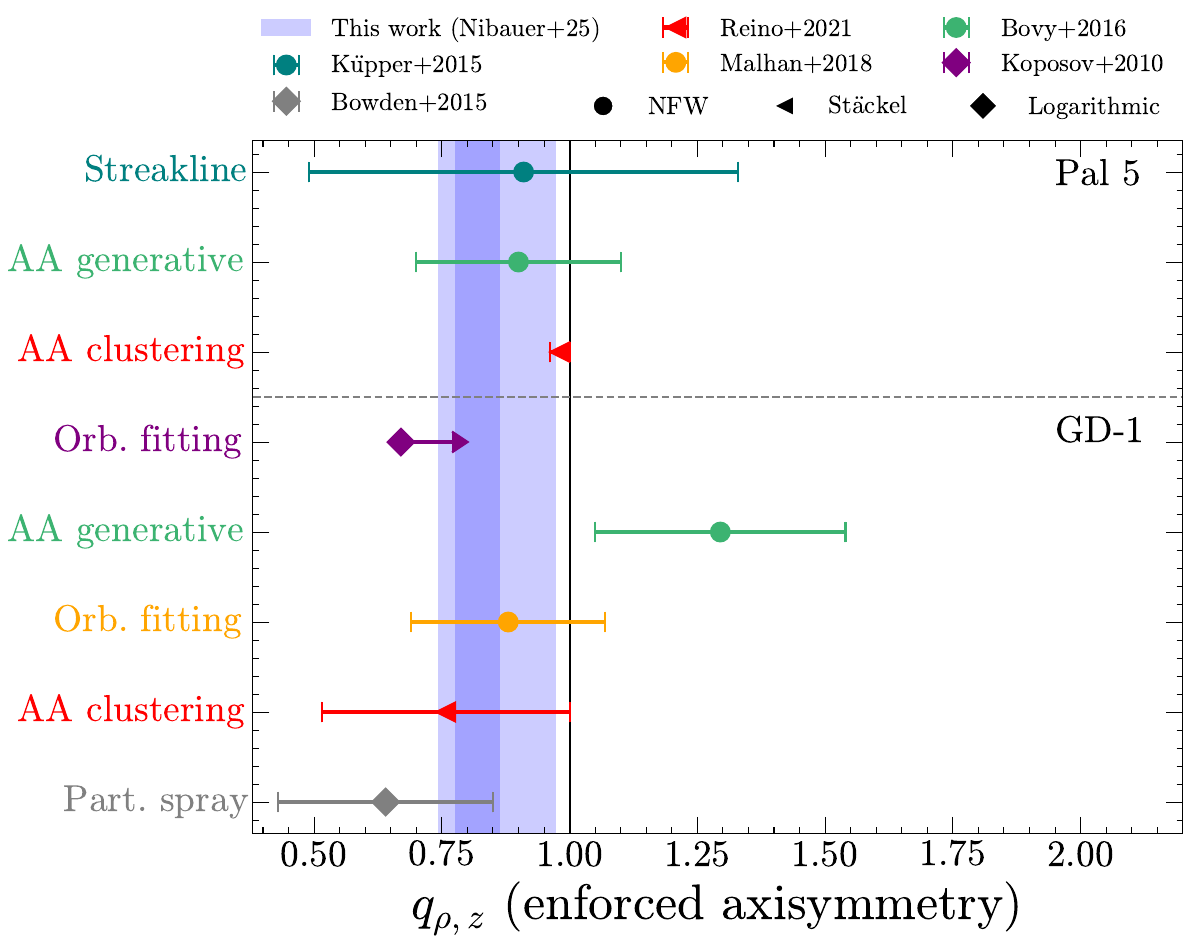}
    \caption{The shape of the inner halo when assuming an axisymmetric global mass model. Our constraint is shown by the purple error-band. Constraints on the $z-$axis flattening in the density are shown for Pal 5 (above the dashed line) and GD-1 (below). Symbols indicate the potential model, while the y-axis labels indicate the method used to model the stream (streakline, generative action-angle, action-angle clustering, orbit fitting, and particle spray).} 
    \label{fig: axisymm_flattening}
\end{figure}

We measure the z-axis halo flattening in density of $q_{\rho, z} = 0.81\pm^{0.06}_{0.03}$ (in \S\ref{sec: results} we reported {\it{total}} flattening in the potential). In Fig.~\ref{fig: axisymm_flattening} we compare our measurement (purple bands for 68 and 95\% confidence) to literature values based on the Pal 5 stream (top) and GD-1 stream (bottom). For works that report potential flattening, we convert to density flattening using $(1-q_\rho)\approx 3(1-q_\Phi)$ (\citealt{2008gady.book.....B}, page 48). The method used to make each measurement is listed next to the data point, along with the reference (color) and potential model (symbol). Most studies adopt a similar mass model to ours, however, \citet{2015MNRAS.449.1391B} report total flattening, instead of halo flattening, and \citet{2021MNRAS.502.4170R} allow for spherical or oblate halos (not prolate) when using a two-component model. For completeness, we include both studies in our comparison.

From Fig.~\ref{fig: axisymm_flattening} there is no clear consensus on the shape of the inner dark matter halo of the Milky Way. While most works prefer an oblate halo, constraints from \citet{2016ApJ...833...31B} imply a prolate halo. Discrepancies may be due to a more limited number of stream members, especially with precise radial velocities and distances, employed in earlier works, which we leave to future work to explore. Another possibility is that an axisymmetric halo model is not the correct functional form, leading to potential systematic errors when attempting a density reconstruction. We explore this possibility in \S\ref{sec: triaxial_halo}. 

\subsection{Triaxial Halo}\label{sec: triaxial_halo}
We now fit a triaxial NFW halo with arbitrary rotation and flattening to the inferred GD-1 accelerations. Free parameters include the halo mass $M$, axis ratios $c/a$, $b/a$, scale-radius $r_s$, pitch, and roll angles. We use the disk, nucleus, and bulge model from \citet{2015ApJS..216...29B}.

\begin{figure*}
\centering\includegraphics[scale=.72]{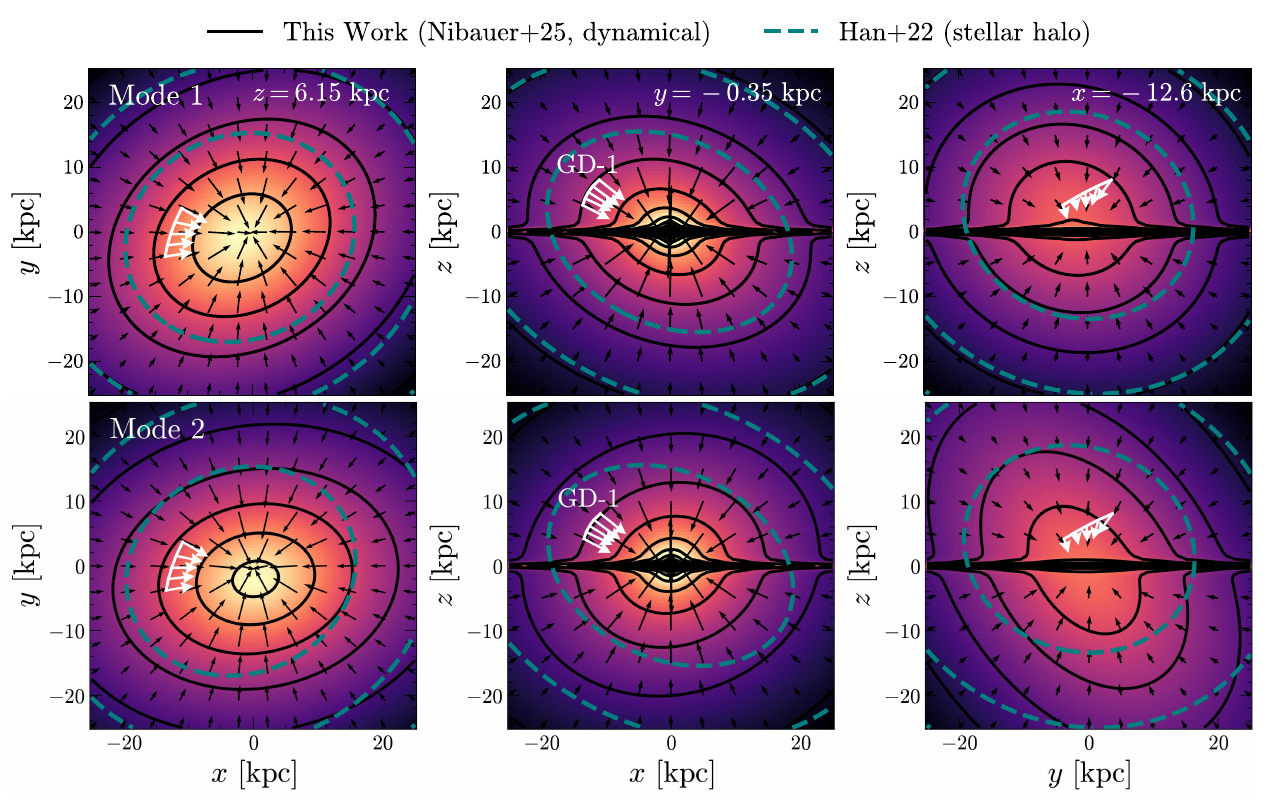}
    \caption{We fit an ellipsoidal density model with rotation to the inferred accelerations, and find two posterior modes that are compatible with our measurement (Mode 1 in top row, Mode 2 in the bottom). The total matter density (heatmap and black contours) of both modes are plotted in Galactocentric slices at the median location of GD-1 (white text labels). White arrows indicate the inferred median accelerations of GD-1, and black arrows correspond to the fitted triaxial model. Mode 1 has density axis ratios of  1:0.75:0.7 and a halo tilt of $\approx 18~\rm{deg}$ out of the disk's plane in the direction of the Sun, remarkably similar to stellar halo constraints (dashed cyan contours, \citealt{2022AJ....164..249H}). Mode 2 has axis ratios 1:0.95:0.65, and a tilt angle of $\approx 56~\rm{deg}$. Acceleration measurements at different locations are needed to distinguish the two modes.}\label{fig: 3d_Best_Fit}
\end{figure*} 

Our sampling of the free parameters reveals two viable configurations for the dark matter halo, labeled as Mode 1 and Mode 2. The inferred global mass distribution is shown in Cartesian coordinates in Fig.~\ref{fig: 3d_Best_Fit} for both modes (1 on top, 2 on bottom). In each Cartesian slice, we evaluate the density (heatmap) and acceleration (black arrows) of the models, and fix the third not plotted dimension to the median location of GD-1 (listed in white text). The median inferred accelerations along the GD-1 stream are shown as white arrows. The first mode prefers a mildly triaxial halo with density axis ratios $c/a =0.70 \pm^{0.06}_{0.04}$, $b/a = 0.75 \pm 0.05$. The pitch angle is $18 \pm^{8}_{5}~\rm{deg}$ and the yaw angle is $23 \pm^{7}_{13}~\rm{deg}$. The second mode, (Mode 2; bottom row) has $c/a = 0.65 \pm 0.05$,  $b/a =0.95 \pm^{0.06}_{0.04}$, pitch angle $56 \pm 8~\rm{deg}$ and yaw angle $97 \pm^{8}_{11}~\rm{deg}$.

For comparison, we overplot stellar halo ellipsoids (dashed blue lines, \citealt{2022AJ....164..249H}). There is a strong agreement between the halo shape implied by our inferred accelerations (Mode 1), and the constraint from \citet{2022AJ....164..249H}, who report $c/a = 0.73 \pm0.02$, $b/a = 0.81\pm 0.03$, pitch angle $25 \pm 3~\rm{deg}$, and yaw angle $24 \pm^6_5~\rm{deg}$. Mode 2 produces similarly compatible local accelerations with our GD-1 inference, though is dissimilar to \citet{2022AJ....164..249H} in the $x-z$ and $y-z$ dimensions. Both modes strongly disfavor alignment of the disk with one of the symmetry planes of the halo.

\section{Summary \& Discussion}\label{sec: summary_and_discuss}
We have measured the Galactic acceleration field in the vicinity of the GD-1 stellar stream. The inferred accelerations are data-driven, and do not rely on analytic models for the gravitational potential. We measure an enclosed mass of $M_{\rm enc}(<14~\rm{kpc}) = 1.41\pm 0.07 \times 10^{11}~M_\odot$, and overall potential flattening of $q_{\Phi} = 0.91 \pm 0.05$, both in good agreement with prior works (e.g., \citealt{2015ApJ...803...80K,2016ApJ...833...31B, 2019MNRAS.485.3296W}). 

We fit our inferred accelerations with two global mass models for the dark matter halo. Assuming axisymmetry, we infer halo  z-axis density flattening of $q_{\rho, z} = 0.81\pm^{0.06}_{0.03}$, consistent with prior works \citep{2015MNRAS.449.1391B,2010ApJ...712..260K, 2019MNRAS.486.2995M, 2016ApJ...833...31B}. However, an axisymmetric model is discrepant with our inferred accelerations at the $2\sigma$ level. A triaxial halo that is tilted with respect to the disk is able to reproduce the full inferred acceleration profile. We find two possible configurations for a tilted dark matter halo. The first (Mode 1) has axis ratios 1:0.75:0.70 (i.e., prolate internal symmetry), with a major axis tilted by roughly $18~\rm{deg}$ with respect to the disk's midplane, and a yaw angle of $23~\rm{deg}$. The second (Mode 2) has axis ratios 1:0.95:0.65 (i.e., oblate internal symmetry) and is tilted by $56~\rm{deg}$ with respect to the disk, with a yaw angle of $97~\rm{deg}$. The first mode is consistent with shape measurements of the stellar halo from \citet{2022AJ....164..249H,2024arXiv240612969H}, and similar to measurements from  \citet{2019MNRAS.482.3868I} who also find a $\sim 20~\rm{deg}$ tilt but with a yaw angle of roughly 70~\rm{deg}. The Mode 1 tilted halo is also able to reproduce the warp and flare of the Galactic disk \citep{2023NatAs...7.1481H}. The second mode is inconsistent with these works. 

Comparing to other dynamical tracers of the halo shape, triaxial axis ratios for the inner halo (within $40~\rm{kpc}$) from the Sagittarius stream are 1:0.97:0.44 \citep{2010ApJ...714..229L}, preferring significantly more $c/a$ flattening than our inference. However, \citet{2010ApJ...714..229L} do not account for the effect of the LMC on the Sagittarius stream, nor allow for an inner halo tilt. \citet{2021MNRAS.501.2279V} find an oblate inner halo when modeling the Sagittarius stream with the LMC, though their inner halo model is disk-aligned by construction. From the phase-mixed Helmi Streams, \citet{2024A&A...691A.277W} measure axis ratios 1:0.84:0.83 (along Galactic X:Y:Z), which is comparable to our Mode 1 constraint. However, their work also assumes a fixed orientation for the principal axes.

Tilted halos are expected outcomes in cosmological simulations both due to the buildup of galaxies from mergers \citep{1989MNRAS.237..785O, 1999ApJ...513L.107D, 2021MNRAS.504.6033S, 2021ApJ...913...36E, 2023ApJ...957L..24H} and due to the instability of the stellar disk when aligned with the principal planes of a triaxial halo \citep{2013MNRAS.434.2971D}. A tilt angle of $18~\rm{deg}$ (Mode 1) is common in IllustrisTNG simulations \citep{2019ComAC...6....2N,10.1093/mnras/stz2338}, with $50\%$ of dark halos tilting $>10~\rm{deg}$, and $25\%$ tilting $> 20~\rm{deg}$. A tilt angle of $56~\rm{deg}$ (Mode 2) is less common, with only $15\%$ of halos tilting $>40~\rm{deg}$ \citep{2023ApJ...957L..24H}. 

The axis ratios we have measured can be characterized with the triaxiality parameter, $T \equiv (1 - (b/a)^2) / (1-(c/a)^2)$. For Mode 1 we find $T = 0.86$, and for Mode 2, $T = 0.17$. A triaxiality parameter of $\approx 0.8$ is expected from dark matter-only simulations, though $T < 0.3$, indicating an oblate mass distribution, is unusual \citep{2006MNRAS.367.1781A,2012JCAP...05..030S}. However, it is likely that baryonic physics drives diversity in halo triaxiality, especially within the inner 30~\rm{kpc} \citep{2008ApJ...681.1076D,2010MNRAS.405.1119K, 2023A&A...669A.132P,2023ApJ...957L..24H}. Even though the tilt and triaxiality of the Mode 1 configuration is preferred in cosmological simulations, Mode 2 is not entirely ruled out, and additional constraints are needed to determine the shape of the Milky Way halo.

In addition, we tested whether non-axisymmetric accelerations can be explained by the LMC or spiral arms. Even though the long axis of the halo tilt is not aligned with the LMC, the LMC is very massive (e.g., \citealt{2021MNRAS.501.2279V}) and may still affect acceleration measurements. For our test, we use $M_{\rm LMC} = 1.5\times10^{11}~M_\odot$, $r_s = 10.8~\rm{kpc}$ \citep{2021MNRAS.501.2279V}. The LMC produces an azimuthal acceleration of $a_{\phi,\rm{LMC}} \approx 0.09~\rm{km \ s^{-1} \ Myr^{-1}}$, whereas we infer larger values of $a_\phi \approx 0.65~\rm{km \ s^{-1} \ Myr^{-1}}$ for $\phi_1 \gtrsim -10~\rm{deg}$. We also tested whether the travel velocity (and acceleration) of the Milky Way disk \citep{2021MNRAS.501.2279V, 2024arXiv240601676C} in combination with the LMC's potential could explain our $a_\phi$ inference. We use the rigid moving potential approach \citep{2022MNRAS.511.2610C}. This experiment gives us $a_{\phi,\rm{LMC}} \approx -0.05~\rm{km \ s^{-1} \ Myr^{-1}}$, which is small in magnitude and of an opposite sign to our inference. Statistically, we find that the LMC's accelerations alone are incompatible with the inferred $a_\phi$ at the $2-3\sigma$ level. Therefore, a tilted halo model provides a significantly better fit than a model with a disk, spherical halo, and LMC. For spiral arms, we tested the potential from \citet{2002ApJS..142..261C} and find that the resulting $a_\phi$ from this model is negligible compared to our inferred accelerations.

While GD-1 is unlikely to have been perturbed significantly by the LMC, the stream does show signs of perturbations due to small-scale structures in the Galaxy (e.g., \citealt{2018ApJ...863L..20P, 2019ApJ...880...38B}). Because our method models the mean phase-space track of the stream rather than its density, we do not expect small-scale perturbations to pose a significant challenge to our methodology, unless perturbations were recent enough to cause a significant track-proper motion misalignment (e.g., \citealt{2019ApJ...885....3S, 2019MNRAS.487.2685E, 2023MNRAS.518..774L, 2023MNRAS.521.4936K}). We find no such evidence of a misalignment, as solar-reflex corrected proper motions are found to point along the stream track. Additionally, local velocity distortions will not be captured by our smooth spline model due to the limited number of knots. 

We have presented the first fully data-driven measurement of the Galactic acceleration field from a stellar stream, and find that 
local accelerations imply that the dark matter halo of the Galaxy is misaligned with the disk. In order to test the robustness of our constraint, additional streams can be independently studied to explore whether our constraints are truly representative of the total inner halo, or only local to the GD-1 stream. Data from precision radial velocity surveys such as DESI \citep{2024AJ....168...58D}, S5 \citep{2019MNRAS.490.3508L}, 4MOST \citep{2019Msngr.175....3D}, WEAVE \citep{2024MNRAS.530.2688J}, and Via (Via collaboration, in prep.), will provide the necessary information to construct a precise 6D phase-space characterization of streams, and directly map the distribution of dark matter in the Galaxy. 

%TC:ignore
\section*{Acknowledgements}
JN is supported by a National Science Foundation Graduate Research Fellowship, Grant No. DGE 2039656. Any opinions, findings, and conclusions or recommendations expressed in this material are those of the author(s) and do not necessarily reflect the views of the National Science Foundation. We are pleased to acknowledge that the work reported on in this paper was substantially performed using the Princeton Research Computing resources at Princeton University which is a consortium of groups led by the Princeton Institute for Computational Science and Engineering (PICSciE) and Office of Information Technology’s Research Computing. We thank Jesse Han, Jenny Greene, Adrian Price-Whelan, Tara Dacunha, and Vedant Chandra for useful discussions regarding the manuscript.
%TC:endignore

\begin{appendix}
    \section{References for Fig.~3}\label{app: fig_refs}
    \citet{2019MNRAS.484.2832V,2019ApJ...875..159E,2022MNRAS.510.2242W, 2020ApJ...894...10L,2020MNRAS.494.4291C,2022ApJ...924..131S, 2022MNRAS.511.2610C, 2021MNRAS.501.2279V,2023MNRAS.521.4936K, 2019ApJ...873..118W,2020MNRAS.494.5178F,2022MNRAS.516..731B,2021MNRAS.502.4170R,2025MNRAS.538.1442L, 2019MNRAS.485.3296W, 2019A&A...621A..56P,2023RAA....23a5013S,2019MNRAS.484.5453C,2021MNRAS.508.5468H,2021MNRAS.501.5964D, 2022ApJ...925....1S, 2019MNRAS.486.2995M, 2019MNRAS.487.2685E,2023MNRAS.524.2124P,2024ApJ...967...89I}
\end{appendix}

\bibliography{thebib}
\end{document}